\documentclass[12pt,letterpaper]{JHEP3}
\usepackage{graphics}
\usepackage{epsfig}

\title{Can the Copernican principle be tested by cosmic neutrino background?}
\author{Junji Jia\\
Department of Applied Mathematics, University of Western Ontario, London, Ontario, N6A 5B7, Canada\\
\email{jjia5@uwo.ca}}
\author{Hongbao Zhang\\
Perimeter
Institute for Theoretical Physics, Waterloo, Ontario, N2L 2Y5,
Canada\\
Department of Astronomy, Beijing Normal University, Beijing,
100875, China\\
\email{hzhang@perimeterinstitute.ca}}

\abstract{The Copernican principle, stating that we do not occupy any special place in our universe, is usually taken for granted in modern cosmology. However recent observational data of supernova indicate that we may live in the under-dense center of our universe, which makes the Copernican principle challenged. It thus becomes urgent and important to test the Copernican principle via cosmological observations. Taking into account that unlike the cosmic photons, the cosmic neutrinos of different energies come from the different places to us along the different worldlines, we here propose cosmic neutrino background as a test of the Copernican principle. It is shown that from the theoretical perspective cosmic neutrino background can allow one to determine whether the Copernican principle is valid or not, but to implement such an observation the larger neutrino detectors are called for.}
\begin{document}

\section{Introduction}
Based on the cosmological principle, the recent observations of type Ia Supernova indicate that either our present universe is dominated by some new exotic matter called dark energy or a modification of general relativity is needed on cosmic scales at least in an effective sense, which raises profound questions on fundamental physics. A conservative way out is to engage in the speculation on the validity of the cosmological principle.
As an assumption of the isotropy and homogeneity throughout our universe, the cosmological principle is known to be partly satisfied. The observation of near-isotropy of cosmic microwave background(CMB) spectrum implies that our universe is very nearly isotropic. In addition, our universe is observed to be approximately homogeneous on large scales\cite{Hogg,YBPS}. However, the radial homogeneity on scales of Gpc remains to be confirmed. In fact, it has been theoretically shown that the spherically symmetric but radially inhomogeneous Lemaitre-Tolman-Bondi(LTB) cosmological models can explain the supernova data very well without introducing the dark energy or resorting to a modification of general relativity\cite{Celerier,Moffat1,Moffat2,Garfinkle,Enqvist,ABNV}. Different from the case of the Friedmann-Robertson-Walker(FRW) cosmological models, in the context of the LTB cosmological models, we are constrained in a special position, i.e., in or near the center of a void where the local matter density is relatively low, which violates the Copernican principle. So whether or not the Copernican principle is valid plays a pivotal role in our understanding of the genuine mechanism underlying the evolution of our universe. Although the Copernican principle may be widely accepted by fiat, it should be observationally tested without an a priori bias.

\begin{figure}[htb!]
\includegraphics[scale=0.5]{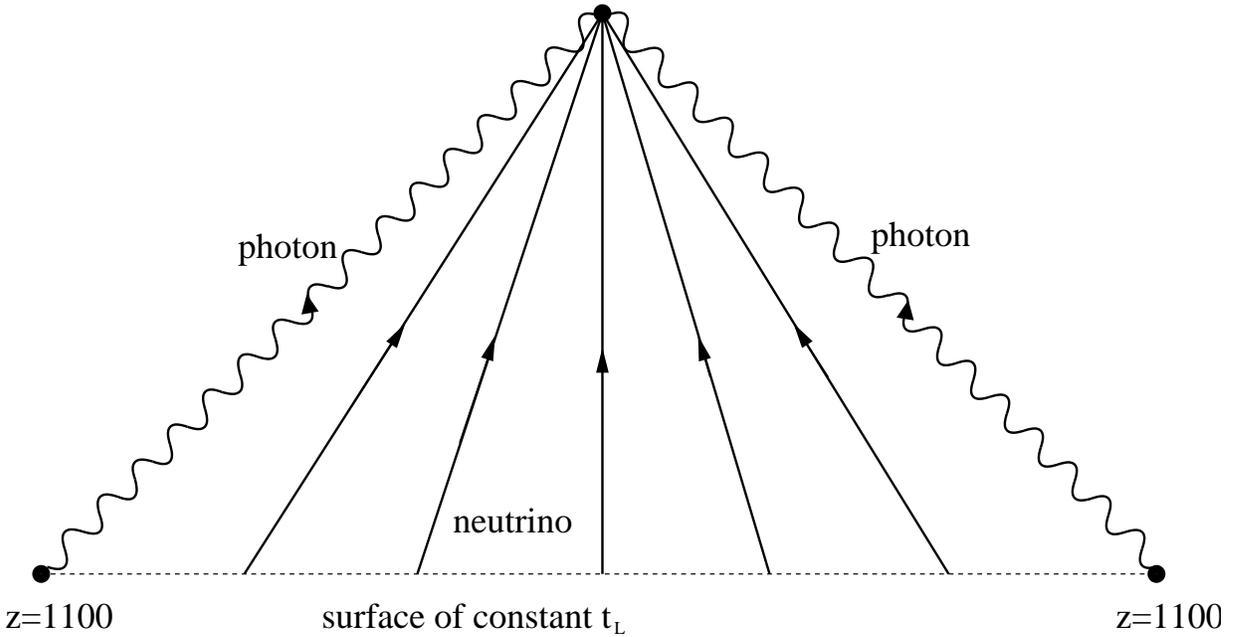}
\caption {\small\sl Different from the cosmic photons, the cosmic neutrinos of different energies come from the different places on the surface of constant $t_L$ and travel to us along the different worldlines.} \label{shooting}
\end{figure}

To date, there have been some ideas proposed to serve as observational tests of the Copernican principle\cite{CS,UCE,CBL,CFL,YKN,BW}. Note that in all of these proposals the observational data come from the past light cone. On the other hand, neutrinos are believed to be massive, however small the mass is. Thus as illustrated in Fig.\ref{shooting}, they will freely travel to us from inside of the past light cone after decoupling, which definitely brings more information about the structure of our universe. Keeping this in mind, we here put forward cosmic neutrino background(CNB) as a new test of the Copernican principle if observed.

In next section, we shall provide a brief review of the LTB cosmological models, where a specific parametrization is also chosen. In subsequent section, we will work out the cosmic neutrino background spectrum in the chosen LTB and FRW cosmological models, and examine the feasibility of our proposal by demonstrating the spectrum difference between the LTB and FRW cases. We then finish the paper with some discussions.

Unless otherwise is explicitly stated, Planck units are used here, i.e., $c=G=\hbar=k=1$.

\section{LTB cosmological models}
Start with the spherically symmetric but radially inhomogeneous LTB metric
\begin{equation}
ds^2=-dt^2+\frac{R'^2(t,r)}{1-\mathcal{K}(r)}dr^2+R^2(t,r)(d\theta^2+\sin^2\theta d\phi^2), \label{LTB}
\end{equation}
where the prime denotes the derivative with respect to $r$. With the energy momentum tensor given by $T_{ab}=\rho(t,r)(dt)_a(dt)_b$, it follows from Einstein equation that
\begin{equation}
\dot{R}^2=\frac{F(r)}{R}-\mathcal{K} \label{Einstein}
\end{equation}
and
\begin{equation}
\rho=\frac{F'}{8\pi R^2R'}, \label{Energy}
\end{equation}
where the dot denotes the derivative with respect to $t$.
For later convenience, we introduce the quantities $a$, $K$, and $\alpha$ as follows
\begin{equation}
R=ar, \mathcal{K}=Kr^2, F=\alpha r^3. \label{Transformation}
\end{equation}
Whence Eq.(\ref{Einstein}) becomes
\begin{equation}
\dot{a}^2=\frac{\alpha}{a}-K. \label{Friedman}
\end{equation}
The corresponding solutions can be obtained as\cite{Celerier}
\begin{eqnarray}
a=\frac{\alpha}{2K}(1-\cosh u), t-t_i=\frac{\alpha}{2(-K)^{3/2}}(\sinh u-u)|&&K<0, \label{Solution}\\
a=(\frac{9\alpha}{4})^{1/3}(t-t_i)^{2/3}|&&K=0,\\
a=\frac{\alpha}{2K}(1-\cos u), t-t_i=\frac{\alpha}{2K^{3/2}}(u-\sin u)| &&K>0,
\end{eqnarray}
where  $t_i$ is generally a function of $r$, referring to the time at which the shell of dust labeled by $r$ has zero area radius. In what follows, we will fix $\alpha=\frac{4}{9}$ once and for all, which can always be achieved by the coordinate freedom to change $r$ to any function of $r$. In addition, we shall restrict ourselves to those models with $t_i=0$, where the big bang is assumed to occur simultaneously at every point. Furthermore, CMB observations suggest that the ratio energy $\Omega=1$ at large distances whereas observations of galaxy clusters suggest that in our neighborhood of the universe $\Omega_M$ is around $0.3$. Note that in the context of the LTB cosmological models, the ratio density is given by\cite{Enqvist}
\begin{equation}
\Omega=\Omega_M=\frac{F}{H_0^2R_0^3}=\frac{\alpha}{H_0^2a_0^3}=(1-\frac{Ka_0}{\alpha})^{-1},
\end{equation}
where the present Hubble rate $H_0=\frac{\dot{R}_0}{R_0}=\frac{\dot{a}_0}{a_0}$. This implies that $K$ should approach zero at large distances and be negative at small distances. Especially, we here follow \cite{Garfinkle} to parameterize $K$ as
\begin{equation}
K=-\frac{1}{1+\beta^2r^2}.
\end{equation}
Whence $a_0$ can be determined by the ratio density at our position as
\begin{equation}
a_0=\alpha[\Omega_M(0)^{-1}-1].
\end{equation}
We then use Eq.(\ref{Solution}) to get the present value of time $t_0$.

As shown in \cite{Garfinkle}, $(\Omega_M(0)=0.2,\beta=5.1)$ can fit the supernova data very well. So in next section we will only consider the LTB cosmological models with the above parameter, and the parameter $(\Omega_M(0)=0.2,\beta=0)$ as well for a comparison since the latter actually reduces to the FRW cosmological model\footnote{Note that in the homogeneous and isotropic universe cosmic neutrino background does not depend on the specific FRW models indeed, which implies that the result for the FRW cosmological model here is the same as that for more realistic FRW cosmological models such as the concordance $\Lambda$CDM model\cite{Weinberg,Zhang}.}.
\section{CNB thermal spectrum}
\begin{figure}[htp!]
\includegraphics[scale=0.6]{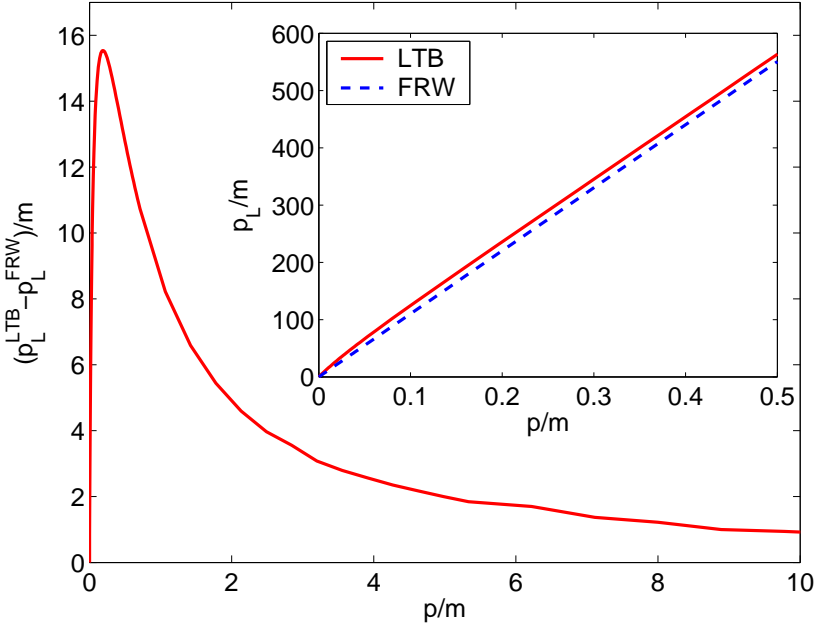}
\caption{\small\sl The $p_L\sim p$ relation for the LTB$(\Omega_M(0)=0.2,\beta=5.1)$ and FRW$(\Omega_M(0)=0.2,\beta=0)$ cosmological models . Here the ratio of momentum of cosmic neutrino to its mass is given by $\sqrt{(\frac{dt}{d\lambda})^2-1}$.} \label{pp}
\end{figure}
Let us first assume the decoupled neutrinos and photons to be both isothermal at the surface of constant $t_L$ where the redshift of photons is $1100$ and the ratio of temperature of neutrinos and photons is $\frac{T_\nu}{T_\gamma}|_{t_L}=(\frac{4}{11})^{1/3}$\cite{Weinberg}. Therefore at the surface of constant $t_L$, the temperature of neutrinos $T_\nu|_{t_L}$ can be obtained by taking into account that the present CMB temperature is $2.7K$, and the number density of neutrinos is given by
\begin{equation}
n(p_L,t_L)=\frac{1}{(2\pi)^3}\frac{1}{\exp(\sqrt{p_L^2+m^2}/T_\nu|_{t_L})+1},
\end{equation}
where $p_L$ represents the momentum of neutrino at the surface of constant $t_L$, and $m$ is the mass of neutrino. We thus be able to work out the present cosmic neutrino background just by propagating the geodesics from us back to the surface of constant $t_L$ and using the collisionless Boltzmann equation, which states that the number density keeps constant along geodesics\cite{Weinberg,EGS}.

With the metric form given by Eq.(\ref{LTB}), the radial geodesics equation can be written as
\begin{eqnarray}
\frac{d^2t}{d\lambda^2}+\frac{R'\dot{R}'}{1-\mathcal{K}}(\frac{dr}{d\lambda})^2&=&0, \nonumber\\
\frac{d^2r}{d\lambda^2}+(\frac{R''}{R'}+\frac{1}{2}\frac{\mathcal{K}'}{1-\mathcal{K}})(\frac{dr}{d\lambda})^2+2\frac{\dot{R}'}{R'}\frac{dt}{d\lambda}\frac{dr}{d\lambda}&=&0,
\end{eqnarray}
where $\lambda$ is the affine parameter with the normalization conditions $-(\frac{dt}{d\lambda})^2+\frac{R'^2}{1-\mathcal{K}}(\frac{dr}{d\lambda})^2=0,-1$ for null and timelike geodesics, respectively. In addition, $\mathcal{K}$, $\mathcal{K}'$, $R'$, $\dot{R}'$, and $R''$ can be obtained by Eq.(\ref{Transformation}) and Eq.(\ref{Solution}) as follows
\begin{eqnarray}
\mathcal{K}&=&-\frac{r^2}{1+\beta^2r^2}, \nonumber\\
\mathcal{K}'&=&-\frac{2r}{(1+\beta^2r^2)^2}, \nonumber\\
R'&=&\frac{\alpha}{2}(\cosh u-1)-\frac{3\alpha\beta^2r^2}{2}(2+\frac{u\sinh u }{1-\cosh u}), \nonumber\\
\dot{R}'&=&\frac{1}{(1+\beta^2r^2)^{3/2}}\frac{\sinh u}{\cosh u-1}+\frac{3\beta^2r^2}{(1+\beta^2r^2)^{3/2}}\frac{\sinh u-u}{(\cosh u-1)^2}, \nonumber\\
R''&=&-3\alpha\beta^2 r(2+\frac{u\sinh u }{1-\cosh u})-\frac{3\alpha\beta^2r}{2(1+\beta^2r^2)}\frac{\sinh u(\sinh u-u)}{\cosh u-1}\nonumber\\
&&-\frac{9\alpha\beta^4r^3}{2(1+\beta^2r^2)}(\frac{\sinh u-u}{\cosh u-1})^2.
\end{eqnarray}
To solve the above radial geodesics equation, we still require the initial conditions, which are apparently obtained by the normalized conditions as
\begin{eqnarray}
\frac{dr}{d\lambda}|_{p_0}>0,&&\frac{dt}{d\lambda}|_{p_0}=-\alpha[\Omega_M(0)^{-1}-1]\frac{dr}{d\lambda}|_{p_0},\nonumber\\
\frac{dr}{d\lambda}|_{p_0}>0,&&\frac{dt}{d\lambda}|_{p_0}=-\sqrt{1+\alpha^2[\Omega_M(0)^{-1}-1]^2(\frac{dr}{d\lambda})^2|_{p_0}},
\end{eqnarray}
for null and timelike geodesics respectively, where $p_0$ is located at $(t_0,0)$, referring to our present position in the universe.

\begin{figure}[htp!]
\includegraphics[scale=0.6]{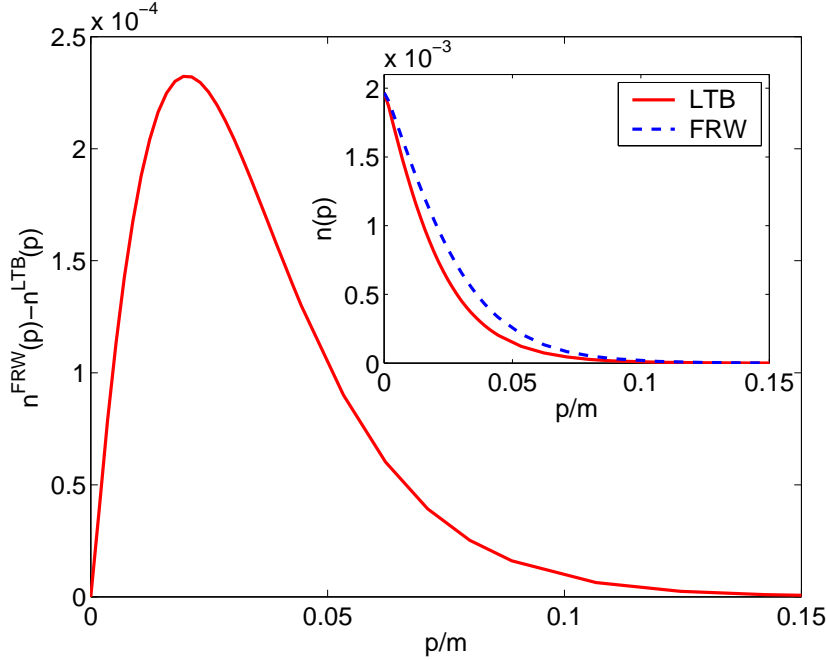}
\caption{\small\sl The $n\sim p$ relation for the LTB$(\Omega_M(0)=0.2,\beta=5.1)$ and FRW$(\Omega_M(0)=0.2,\beta=0)$ cosmological models. Here the present number density of cosmic neutrinos is obtained by the collisionless Boltzmann equation $n(p)=n(p_L,t_L)$, with the neutrino mass assumed to be $0.009ev$\cite{SB}.} \label{np}
\end{figure}

With the data and procedures mentioned above, we can first obtain how the present momentum of neutrino $p$ is related to its momentum $p_L$ at the surface of constant $t_L$ by solving the geodesics equation numerically, which is demonstrated in Fig.\ref{pp}. One can
see that for the FRW cosmological model, as expected $p_L$ is linearly connected with $p$\cite{Weinberg,Zhang}. While for the LTB cosmological model, the $p_L\sim p$ relation has a deviation from the FRW case at relatively small momenta with the peak located at one eighth of neutrino mass or so, and asymptotically approaches the FRW case as the momentum increases, which implies that the effect induced by the inhomogeneity distinctly occurs at a small scale around us and is suppressed at large scales, since as illustrated in Fig.\ref{shooting}, the larger momentum the observed cosmic neutrinos have, the farther distance they travel to us from. This result thus agrees with the primordial motivation for the LTB cosmological models to explain the low redshift supernova data without observable deviation from the approximate homogeneity on large scales.

\begin{figure}[htp!]
\includegraphics[scale=0.6]{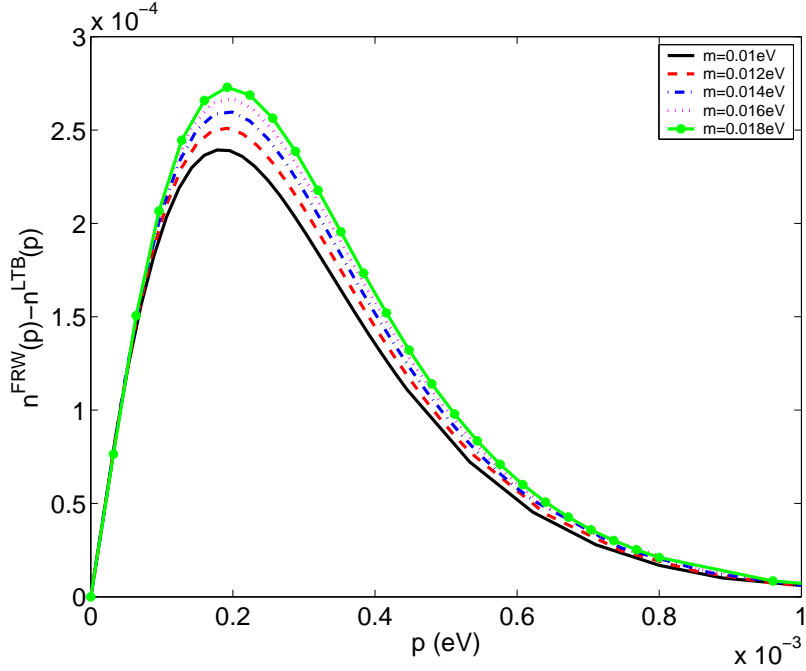}
\caption{\small\sl The mass effect on the difference of $n\sim p$ relation between the LTB$(\Omega_M(0)=0.2,\beta=5.1)$ and FRW$(\Omega_M(0)=0.2,\beta=0)$ cosmological models.} \label{npmass}
\end{figure}

Next simply by the collisionless Boltzmann equation, the $p_L\sim p$ relation yields the present number density of cosmic neutrinos, which is plotted in Fig.\ref{np}. The good news is that the amplitude of difference of number density between the LTB and FRW
cases is comparable to cosmic neutrino background itself, which makes it possible to test the Copernican principle by cosmic neutrino background from the purely theoretical viewpoint. While the bad news is that the peak of difference is situated at a very small momentum, which calls for the instrumentation of very large detection systems since detection cross sections are smaller at lower energies. As demonstrated in Fig.\ref{npmass}, the situation does not change much with the increase of neutrino mass although the larger the mass of neutrino is, the more distinct the signal becomes.  Therefore as a practical matter, it seems unavailable to implement our proposal in our current neutrino detectors.
\section{Discussions}
Motivated by the intuitive observation that the cosmic neutrinos of different energies travel to us from the different places, we have investigated the feasibility of cosmic neutrino background as a test of the Copernican principle. As a result, in the region of small momenta, cosmic neutrino background shows us the definite signal to determine whether our universe is homogeneous or not. However, because this signal shows up at low energies, it seems invisible in our current neutrino detectors. A more detailed evaluation of the possibility to carry out our proposal in the future neutrino telescopes is worthy of further investigation, which goes beyond the scope of this paper. But we expect that our theoretical analysis here manifests the very advantage of neutrinos over photons in bringing extra information about our universe, thus provides another stimulus to such endeavors of construction of large neutrino telescopes.

We conclude with some caveats. Note that our result of $n\sim p$ relation depends on the isothermal distribution assumption of cosmic neutrinos at the time $t_L$. So a deviation from the assumed distribution will enhance or suppress the signal. In addition, so far our discussions have been restricted to the specific LTB cosmological model. More general LTB cosmological models may give us somewhat different results. Although a more realistic neutrino distribution is believed to make the signal enhanced, and the results are expected to display the same qualitative behavior for other viable LTB cosmological models, a careful investigation is needed. Last but not least, if neutrinos are heavy enough, then the local gravitational clustering of cosmic neutrinos will become significant such that the $n\sim p$ relation will be distorted\cite{RW}. It is thus important to see how our result is influenced by this clustering effect. We expect to report these subtle issues elsewhere.
\acknowledgments
HZ is indebted to Song He and Federico Piazza for their
constant encouragements and stimulating discussions during this work. He would also like to thank David Garfinkle for helpful correspondence on the LTB cosmological models. JJ was supported by NSERC of Canada and Ontario graduate scholarship. HZ was supported
by the Government of China through CSC(no.2007102530). This
research was supported by Perimeter Institute for Theoretical
Physics. Research at Perimeter Institute is supported by the
Government of Canada through IC and by the Province of Ontario
through MRI.

\end{document}